\documentclass[useAMS,usenatbib,usegraphicx]{mn2e}

\title[Inverse Compton X--rays from 3C 219]
 {Inverse Compton X--rays from the radio galaxy 3C 219}
\author[A. Comastri et al.]
  {A.~Comastri,$^1$
  G.~Brunetti,$^2$ D.~Dallacasa,$^3$ M.~Bondi,$^2$
  M.~Pedani,$^4$ G.~Setti$^3$\\
  $^1$INAF -- Osservatorio Astronomico di Bologna,  
    via Ranzani 1, I--40127 Bologna, Italy\\
  $^2$Istituto di Radioastronomia del CNR 
 via Gobetti 101, I--40129, Bologna, Italy\\
    $^3$Dipartimento di Astronomia, Universit\`a di Bologna 
    via Ranzani 1, I--40127 Bologna, Italy\\
    $^4$ INAF -- Centro Galileo Galilei -- S/C La Palma, 38700 TF Spain}
\date{Released 2002 December 31}

\pagerange{\pageref{firstpage}--\pageref{lastpage}} \pubyear{2002}

\def\LaTeX{L\kern-.36em\raise.3ex\hbox{a}\kern-.15em
    T\kern-.1667em\lower.7ex\hbox{E}\kern-.125emX}

\begin{document}

\label{firstpage}

\maketitle

\begin{abstract}

We report the results from a {\it Chandra} observation of the powerful
nearby (z=0.1744) radio galaxy 3C 219. We find evidence for non--thermal
X--ray emission from the radio lobes which fits fairly well with a 
combination of inverse Compton scattering of Cosmic Microwave 
Background radiation and of nuclear photons with the relativistic
electrons in the lobes.
The comparison between radio synchrotron and IC emission yields
a magnetic field strength significantly lower ($\sim 3$) than
that calculated under minimum energy conditions; the source  
energetics is then dominated by the relativistic particles.

\end{abstract}


\section{Introduction}

It is well known that diffuse non--thermal X--ray emission
from the radio lobes 
is a unique tool to constrain the spectrum and the energetics
of the relativistic electrons  
and possibly the acceleration mechanisms at work
in radio galaxies.

Diffuse  X--rays from 
the lobes of radio galaxies and quasars
are produced by inverse Compton (hereinafter IC) 
scattering of Cosmic Microwave Background (CMB) 
photons (e.g. Harris \& Grindlay 1979),
and/or nuclear photons (Brunetti, Setti \& Comastri 1997).
In the first case they are sampling the relativistic
electrons with Lorentz factor $\gamma \sim$ 10$^3$, while 
the X--rays from IC scattering of
the nuclear far--IR/optical photons are mainly powered
by $\gamma\sim 100-300$ electrons
whose synchrotron emission typically falls
in the undetected hundred kHz frequency range.

Non--thermal 
X--ray emission from the radio lobes has been discovered by {\tt ROSAT} and
{\tt ASCA} in a few nearby radio galaxies, namely Fornax A (Feigelson et
al. 1995;
Kaneda et al. 1995; Tashiro et al. 2001), 
Cen B (Tashiro et al. 1998), 3C 219 (Brunetti et al. 1999)
and NGC 612 (Tashiro et al. 2000).
The analysis of the spatially coincident X--ray and
radio flux densities has allowed to estimate 
magnetic field strengths which are close, but on average lower 
than the values derived under equipartition conditions.
However these results are affected by significant uncertainties 
arising from the low X--ray counting statistics in the diffuse component
and the insufficient angular resolution of the instruments.

The superb capabilities of the {\it Chandra} X--ray observatory
make now possible to image, on arcsec scale,
the radio lobes of powerful radio galaxies and quasars
and to disentangle the non--thermal emission from the
thermal and from the nuclear component.
Non--thermal emission from the radio lobes of 
relatively compact and powerful objects
has been successfully detected with  
{\it Chandra} in the case of the radio galaxy
3C 295 (Brunetti et al., 2001) 
and, possibly, in a few high redshift radio galaxies (Carilli 2002),
and in the counter lobes of the radio loud quasars 
3C 207 (Brunetti et al. 2002) and 3C 179 (Sambruna et al. 2002).
These observations are well interpreted as IC scattering of 
IR photons from the corresponding nuclear source,  thus
providing a clear evidence for the presence of 
low energy electrons ($\gamma \sim$ 100) in the radio volumes.
On the other hand, the detection of X--ray emission from the
radio lobes of giant radio galaxies is best explained by 
IC scattering of the CMB photons (Hardcastle et al. 2002; 
Isobe et al. 2002; Grandi et al. 2002). 

In this paper we report  on the {\it Chandra} observation of
the nearby, powerful, double--lobed FRII radio galaxy 3C 219.
$H_0=50$ km s$^{-1}$ Mpc$^{-1}$ and $q_0=0.5$ are used throughout.

\section{3C 219 and the radio data}

The radio galaxy 3C 219 at $z$=0.1744 is a prototype of the FR--II class
of powerful radio sources. Its projected angular size is about 180
arcsec (Clarke et al. 1992) corresponding to a projected liner size of
about 680 kpc. The optical counterpart of the nuclear radio source is a
cD galaxy with M$_{V}=-21.4$ (Taylor et al. 1996) in a non-Abell
cluster (Burbidge \& Crowne 1979).
The radio structure is characterized by a relatively weak core, a
straight jet $\sim 20^{\prime\prime}$ long (75 kpc) pointing towards the
SW, and a weak blob on the counterjet side. Two almost symmetric 
hot-spots are present, the one in the south--west lobe being aligned 
with the jet axis.
The two lobes account for most of the radio
emission at all frequencies below 15 GHz. \\
The overall radio spectrum is best fitted with a  
slope $\alpha \sim$~0.8 (S$\sim \nu^{-\alpha}$)
between 178 and 750 MHz (Laing et al. 1983) 
which steepens to $\alpha\simeq$ 1 
above about 1 GHz (see Fig.~3). \\
By assuming the standard equipartition recipe,
which corresponds to integrate the synchrotron spectrum 
between  the fixed range of frequencies 10 MHz -- 100 GHz (Pacholczyk 1970),
the magnetic field in
the lobes, considered as homogeneous emitting regions, is
$\sim$ 7.6 $\mu$G. If we assume instead that low energy electrons
down to $\gamma_{min}\sim 50$ are present, the magnetic field increases
to $\sim$ 10.4 $\mu$G.

\subsection{ The VLA observations}

The radio images at 1.4 GHz were obtained by combining archival VLA data:
A-configuration observations were carried out on May 18, 1986 for a
total integration of 6.5 hours, while B-configuration data were taken
on September 6, 1986 for a total of about 7 hours on source. The total
bandwidth is 25 MHz for each dataset. We carried out the data
reduction in a standard way by means of the NRAO Astronomical Image
Processing System (AIPS). 
We first processed the B-configuration data, and then used the final
image to start the phase self-calibration of the shortest spacings of
the A-configuration data for which the source structure is
resolved out in the latter dataset. 
Then we combined the A and B configuration observations and the final
image obtained has a resolution of 1.74~$\times$~1.64 arcsec in
p.a. $-13^\circ$; the off-source r.m.s. noise level in the image plane
is 40 $\mu$Jy/beam. The total flux density accounted for 3C219 in our
image is 8.525 Jy. \\
A more detailed analysis of the radio data will be presented
elsewhere, where further VLA observations at various frequencies and
configurations will be discussed (Dallacasa et al. in preparation).

\section{X--ray observations}

\subsection{Imaging}

The target was placed at about 40'' from the nominal aimpoint
of the  back illuminated ACIS S3 chip onboard {\it Chandra}
and observed on 2000 October 11th.
Although the observation was performed using a subarray
configuration the nuclear source is affected by photon pile-up
and will not be considered any further.
The raw level 1 data were re--processed using the latest
version 2.2 of the CXCDS {\tt CIAO} software and filtered 
using a standard grade selection.
There is evidence for an increased background count rate towards 
the end of the observation. The corresponding time intervals  
were filtered out leaving about 16.8 ksec of useful data.
In order to study the morphology of the extended emission 
all the counts within a region of 5 arcsec radius centered on 
the nuclear source and those in the read--out streaks have been subtracted.
The full band (0.5--7 keV) X--ray image smoothed with 
a fixed size gaussian (FWHM of 10$^{\prime\prime}$)
is reported in Fig.~1. The radio contours obtained from the  
1.4 GHz radio image smoothed on the same scale are 
also overplotted.

\begin{figure}
\includegraphics[width=84mm]{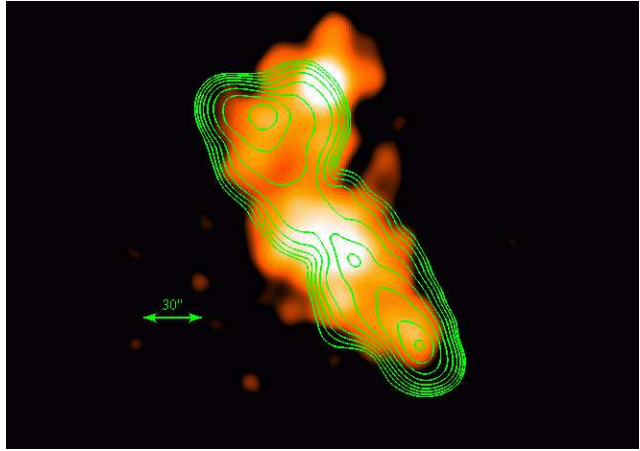}
 \caption{The 0.5--7 keV  X--ray image smoothed with a Gaussian kernel
(FWHM of 10 arcsec). The color scale ranges from about 2$\sigma$ up 
to 9$\sigma$ above the average local background level.    
The nuclear component and the pile--up streaks have been 
subtracted before smoothing. 
The overlayed radio--contours at 1.4 GHz have been obtained from 
the VLA image smoothed with the same Gaussian kernel. Contour peak 
radio flux is at 0.64 Jy/beam and contours level are spaced by a factor 
of 2, the lowest contour being 0.0025 Jy/beam.}
 \label{sample-figure}
\end{figure}

The detection of extended X--ray emission 
almost coincident with the radio lobes emission 
strongly favour the non--thermal IC nature for the
high energy radiation as previously suggested 
by Brunetti et al. (1999) on the basis of lower resolution ROSAT HRI data. 

The bright clump at the boundary of the northern radio lobe is characterized
by a soft X--ray spectrum and is most likely due 
to a previously unknow group of galaxies at z$\sim$ 0.4 
(see section 3.2.3).
The relatively faint diffuse emission coincident with the radio 
lobes shows a brightness increment in the innermost part of the
northern lobe. Extended X--ray emission to the east 
from the nucleus not spatially coincident with the radio contours
is also evident (Fig.~1).
Finally, there is also a jetted X--ray emission 
at about 10--30 arcsec south of the nucleus which appears
to be spatially coincident with
the two radio knots of the main jet (Fig.~2).

\begin{figure}
\includegraphics[width=84mm]{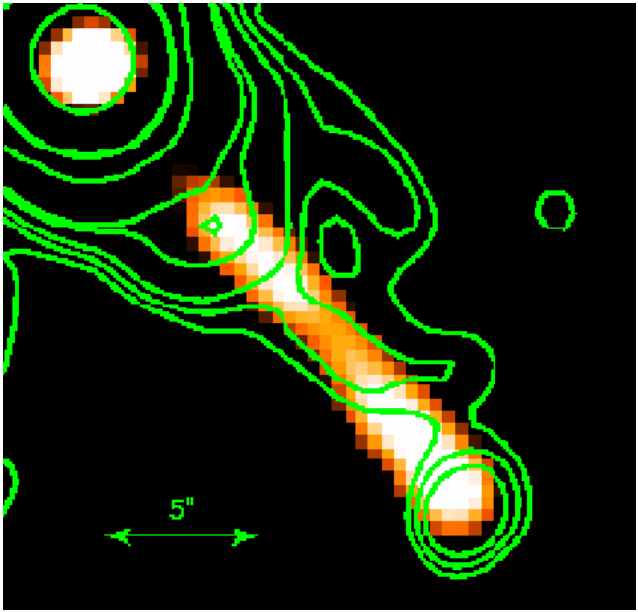}
 \caption{An enlarged view of the jet and nuclear region. The colour scale
corresponds to the 1.4 GHz radio image smoothed with  
2.5$^{\prime\prime}$ resolution (FWHM) 
The overlayed countours have been obtained from the hard band (1.5--7 keV) 
X--ray image smoothed with the same scale. 
The X--ray contours correspond to (1.2,1.8,2.4,3,3.6,7.2,15,59,500) counts 
per pixel. The lowest contour is 
about 2$\sigma$ above the local background.}
 \label{sample-figure}
\end{figure}

\subsection{Spectroscopy}

Given that we are interested in the study of the origin of the faint
diffuse X--ray emission from the radio lobes the background subtraction 
is an important issue. 

In order to obtain the best estimate of the total background 
we rely on a technique 
which allows to significantly reduce the particle background 
for those observations carried out in the Very Faint Mode. 
\footnote{${\rm http://cxc.harvard.edu/cal/Links/Acis}$ ${\rm /acis/Cal_prods
/vfbkgrnd/index.html}$}
We have employed this technique (hereinafter VF) with the recommended 
choice of parameters. As a result the
quiescent background is reduced up to 30\% depending on the 
energy range.
Source spectra have been extracted
using appropriate response and effective area functions taking into
account the source extension and weighing the detector response 
and effective area according to the source spectrum.
The instrument response has been also corrected for the 
degradation in the {\tt ACIS} quantum efficiency using the latest version 
of the {\tt ACISABS} tool.  

\subsubsection{The lobes spectra}

The {\it Chandra} spectrum of the entire extended emission (Fig.~3), 
including the jet knot, but excluding the background cluster (see below),  
is best fitted with a power law with $\Gamma=1.74\pm0.17$
plus Galactic absorption ($N_H = 1.55 \times 10^{20}$ cm$^{-2}$)
fully consistent with the radio spectrum.
The 0.5--7 keV flux of  2.4 $\times$ 10$^{-13}$ erg cm$^{-2}$ s$^{-1}$ 
corresponds
to a rest--frame luminosity of about 3 $\times$ 10$^{43}$ erg s$^{-1}$.
Even though the counting statistics (about 570 net counts) does not allow to 
statistically discriminate between a power law and a thermal model 
the latter returns a slightly worse fit ($\Delta\chi^2 \simeq$ 3 for the
same number of parameters) with a 
best fit temperature of about 4 keV (greater than 3 keV
at the 90\% confidence level). Besides the fact that the
observed 0.5--7 keV luminosity is about an order of magnitude lower 
than that expected from the $L_X$--T relation for clusters of galaxies
(e.g., Arnaud \& Evrard 1999), 
the observed morphology and the extremely 
good agreeement between the X--ray and radio spectral slopes 
strongly rules out a thermal origin for the diffuse X--rays.

In order to search for possible spectral differences between the 
northern and southern lobe (excluding the jet region)  
we have considered the X--ray spectra of 
the two regions separately. We find that the two regions have approximately 
the same spectral shape, which now is rather poorly constrained 
with a marginal indication of a slightly harder spectrum 
($\Delta\alpha \simeq$ 0.25) in the southern lobe.

\begin{figure}
\includegraphics[angle=-90,width=84mm]{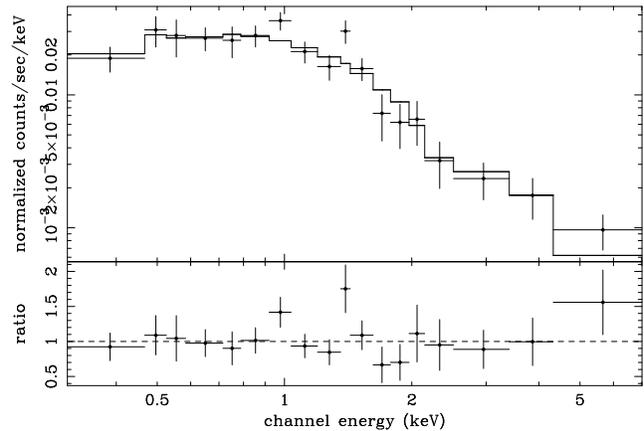}
 \caption{The 0.3--7 keV spectrum of the diffuse emission fitted with 
a power law model plus Galactic absoprtion.}
 \label{sample-figure}
\end{figure}

\subsubsection{The jet knot}

There is significant evidence of an excess of X--ray emission 
in the region coincident with the southern 
radio jet and in particular with the brightest knot (Fig.~2). 
Although only 12 net counts are detected in the 0.5--7 keV
energy range, the band ratio H/S $\simeq$ 2 (where H are the counts in the 
2--7 keV and S in the 0.5--2 keV bands, respectively) folded 
through the  ACIS--S spectral response corresponds to a power law 
spectrum with $\Gamma\simeq$ 0 or to an 
intrinsic absorption column density of about 10$^{22}$ cm$^{-2}$
if  $\Gamma=1.7$ is assumed.
 However taking into account the errors associated with 
small numbers statistic (Gehrels 1986) the spectral slope could be 
as steep as $\Gamma \simeq 2$ (90 \% confidence) and the absorption
fully consistent with the Galactic value.    
A deeper exposure of 3C219 is needed to confirm the hardness
of the X--ray jet and thus constrain the emission mechanism.

\subsubsection{The background cluster}

A R--band observation retrieved from the HST archive shows the presence of
an excess of optical galaxy counts at the position of the north west clump,
possibly indicating the presence of a cluster/group.
We have obtained the spectrum of the most luminous galaxy of
the group with the 3.5m Telescopio Nazionale Galileo 
(TNG) and identify it with an elliptical galaxy at $z$=0.389.
The X--ray spectrum of the north west clump (about 115 net counts) 
is well fitted by a thermal model with a relatively low 
temperature, $kT \sim 2.1$ keV plus Galactic absorption.
The 0.5--2 keV flux is about 2 $\times 10^{-14}$ erg cm$^{-2}$ s$^{-1}$
and the corresponding luminosity is about 1.2 $\times 10^{43}$ erg s$^{-1}$
which is within a factor of $\sim 2$ of that expected from the 
$L_X$--T relationship.

\section{Discussion}

The morphology and scale of the faint diffuse X--ray emission is very 
similar to that of the radio lobes
thus suggesting a non--thermal mechanism due to
IC scattering of CMB and nuclear photons.
The latter would dominate at distances from the nucleus:

\begin{equation}
R_{\rm kpc} <
70 \times L_{46}
 \left( 1 - \mu \right)^2 \left( 1 + z \right)^{-2}
 \label{conf}
\end{equation}

where $L_{46}$ is the isotropic nuclear luminosity in units of $10^{46}$
erg s$^{-1}$  
in the far--IR to optical band and $\mu$ is the cosine of the
angle between the direction of the nuclear seed
photons and the scattered photons (approximated with the 
angle between radio axis and line of sight
with $\mu$ negative for the far lobe and positive for 
the near one).
Since we detect diffuse X--ray emission coincident with the
radio lobes out to a distance of $\sim$300 kpc from the nucleus
with a roughly constant and symmetric brightness (Fig.~1),
we conclude that the IC scattering of CMB photons is the
process which accounts for the majority of the observed X--rays.
In addition, we note the presence of enhanced X--ray brightness
in the innermost $\sim$ 15--20 arcsec (projected size $\sim$ 
70 kpc) of the northern (counter) lobe.
Although the modest number of counts in this region does not 
allow to prove the non--thermal origin of this excess, we point 
out that it would be naturally explained by an additional 
contribution from  IC scattering of nuclear photons which, 
indeed, is expected to be stronger in the counter lobe 
(Brunetti et al. 1997). 
In particular, assuming an angle between the radio axis and the 
plane of the sky of 30$^o$ (appropriate for one--sided jet 
FR II radio galaxies), Eq.(\ref{conf}) yields a far--IR to 
optical luminosity of the hidden quasar of 
$\sim 6 \times 10^{45}$ erg s$^{-1}$ which is only a factor of 
$\sim 2$ lower than that estimated by Brunetti et al. (1999) by
making use of a quasar SED normalized to the nuclear X--ray flux.
We stress that the present estimate does
not depend on the energy densities of the electrons and of the
magnetic field in the radio lobes.

From Fig.~1 the X--ray brightness distribution
of the enhanced central region appears to be somewhat inclined 
toward north east with respect to the radio
axis. Although we cannot exclude that thermal emission could 
be responsible for a fraction
of the X--ray flux in this region, the observed misalignement indicates 
an offset between the radio axis
and the axis of the pattern of the nuclear photons.
We also note that this emission extends 
further on toward east than the radio pattern, while the diffuse  
X--ray emission in the lobes is generally
well confined within the radio isophotes. 
This effect finds a straightforward explanation 
in the  model of IC scattering of the nuclear photons 
because the X--rays from IC scattering of the nuclear photons
are emitted by $\gamma \sim$ 100-300
electrons which have radiative life--times at least
10 times longer (depending on the importance of Coulomb losses)
than the $\sim$ GHz radio emitting electrons.
The oldest population of relativistic electrons in the radio 
lobes may not be visible anymore in the 1.4 GHz image but still
sampled by the {\it Chandra} image.
As an indication in favour of this possibility it should
be noted that the 74 MHz radio observations, which sample
electrons with lower energies ($\gamma\sim$ 2000)
than those synchrotron emitting at 1.4 GHz, also show a slightly
larger extension of the radio 
lobes in the same region with respect to the higher frequency
observations (Blundell, Kassim \& Perley 2000). 
In addition, we note that the innermost radio contours in the northern lobe
are asymmetric with respect to the radio axis
and that they seem to trace a back flow which point
in the direction of the X--ray brightness excess region.

\begin{figure}
\includegraphics[width=84mm]{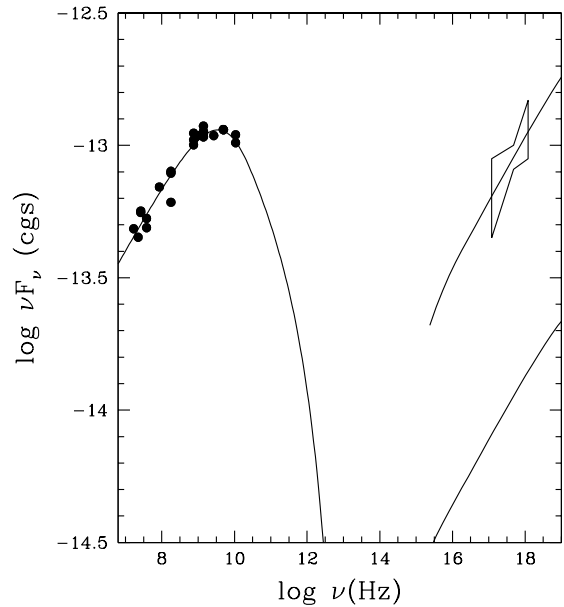}
 \caption{The broad band energy distribution of 3C 219. The radio points 
have been compiled from the NED. The bow--tie represents 
the 90\% confidence countours on the X--ray spectral shape 
from {\it Chandra} data. The radio data are fitted with synchrotron 
emission from a distribution of continuously injected 
relativistic electrons without a low energy flattening. 
The IC emission expected 
from scattering of CMB photons are reported 
assuming equipartion (lower curve) and the magnetic field 
required to match the X--ray flux (see text).}
 \label{fig4}
\end{figure}

The lobes broad band emission (Fig.~4) has been computed by 
assuming a relativistic electron spectrum with a slope
$\delta=2.5$. The corresponding   
synchrotron break ($\Delta \delta =1$) and high frequency 
cut--off are constrained by the radio data at 10$^{10}$ and 
$\geq$10$^{13}$ Hz respectively. The observed X--ray emission is 
accounted for by IC scattering of CMB photons
if $B \simeq 3.1 \mu$G. Such a value is a factor $\sim 2.5$ lower
than that would be obtained with the standard 
equipartition formulae (e.g., Packolczyk, 1970)
and about a factor 3.3 lower if the equipartition
formulae are based on a reasonable choice 
\footnote{We assume a ratio between protons and electron 
energy density $k=1$ and a filling factor $\phi=1$}
of the low energy cut--off in the electron spectrum
($\gamma_{\rm low}=50$; Brunetti et al. 1997).
We have also tried to measure the magnetic field across the 
source by making use of the synchrotron and IC fluxes of both
the southern and northern radio lobes.
We obtain $B \sim 2.9 \mu$G and $B \sim 3.6 \mu$G for the 
northern and southern lobe, respectively.
These values imply that, depending on the adopted 
equipartition formulae, the electron energy density is 
a factor 40--100 and 20--70 larger than that of the average 
magnetic field in the northern and southern radio lobes, 
respectively.
We remind that the equipartition magnetic field strength 
would increase if the assumption on the $\phi$ and $k$ values are relaxed.
These results confirm our previous findings (Brunetti et al. 1999)
reached on the basis of {\tt ROSAT} and {\tt ASCA} data.

The departure from equipartition of 3C 219 is similar 
to that found in the lobes of other giant radio galaxies and quasars which  
emit X--rays via IC of CMB photons (e.g., Cen B, Tashiro et al. 1998;  
3C 351, Hardcastle et al. 2002; 3C 452, Isobe et al. 2002; Pictor A,  
Grandi et al. 2003).   
On the other hand, despite the poor statistics, the lobes of smaller 
radio sources, where the effects of IC scattering of the nuclear 
photons dominates, seem to be closer to minimum energy 
conditions (e.g., 3C 295, 3C 207; Brunetti et al. 2001,2002). 
Finally the X--ray emission of 3C 219 does not appear 
to extend all the way to the end of the radio lobes (Fig.~1),  
possibly suggesting an amplification of the magnetic field in these 
external regions as also found in other IC/CMB radio galaxies  
(e.g., Cen B, Tashiro et al. 1998; 3C 452, Isobe et al. 2002).

\section{Conclusions}

The relatively short (16.8 ksec) {\it Chandra} observation of 
the radio galaxy 3C 219 has successfully detected diffuse 
X--ray emission from the radio lobes and X--rays associated 
with the main radio jet.
The morphology, extension and spectrum of the diffuse
emission strongly support a non--thermal origin, most likely
IC scattering of CMB photons.
In addition, the net brightness increment observed in the 
innermost $\sim 70$ kpc of the northern radio lobe (the far lobe)
suggests an additional contribution from IC scattering of nuclear
photons for a reasonable far--IR to optical isotropic luminosity 
of the hidden quasar of $\sim 6 \times 10^{45}$ erg s$^{-1}$. 
This implies that the electron spectrum extends to much lower energies 
(at least $\gamma\sim$ 100) than sampled by the radio emission. 

From the distribution of the central X--ray emission we deduce 
that the radiation pattern of the hidden quasar (molecular 
torus axis) makes an angle of several tens degrees with the radio
axis and that the oldes population of the relativistic 
electrons may occupy a much larger volume than indicated
by the radio contours.

By comparing the radio synchrotron and IC emissions 
from the lobes we obtain an average magnetic field 
strength approximately
a factor $\sim 3$ lower than the equipartition value.

Additional data from a deeper {\it Chandra} observation would 
allow  to perform detailed, spatially resolved X--ray 
spectroscopy of the  diffuse emission and to compute 
$B$-field intensity structure 
and relativistic electrons' distribution, 
if combined with the available radio data.

Finally, the relatively bright X--ray clump at north west on 
the border of the northern radio lobe (Fig.1) is well fitted 
by thermal emission from a relatively cold plasma ($\sim 2$ keV) 
in a cluster/group of galaxies at $z=0.389$.

\section{Acknowledgments}

This research has made use of the NASA/IPAC Extragalactic Database (NED)
which is operated by the Jet Propulsion Laboratory, California Institute 
of Technology, under contract with the National Aeronautics and Space 
Administration.  This letter is based on observations made with 
the Italian national telescope Galileo Galilei (TNG).
We thank Cristian Vignali for a careful reading of the manuscript 
and the anonymous referee for the useful comments and suggestions.      
The authors acknowledge partial support by the ASI contracts 
I/R/113/01 and I/R/073/01 and the MURST grant Cofin--01--02--8773.

%
%

%

\end{document}